\documentclass[twoside,12pt]{article}
\input{epsf.tex}
\usepackage[dvips]{graphics,epsfig,graphicx}
\usepackage[latin1]{inputenc}
\usepackage{textcomp}
\def\laq{\raise 0.4 ex \hbox{$<$}\kern -0.8 em\lower 0.62 ex\hbox{$\sim$}}
\def\gaq{\raise 0.4 ex \hbox{$>$}\kern -0.7 em\lower 0.62 ex\hbox{$\sim$}}

\def\beq{\begin{equation}}
\def\eeq{\end{equation}}
\def\beqa{\begin{eqnarray}} 
\def\eeqa{\end{eqnarray}}

\begin{document}
\pagestyle{plain}

\begin{flushright}
DF/IST--8.2009\\
December 23, 2009
\end{flushright}
\vspace{15mm}

\begin{center}

{\Large\bf Towards a Noncommutative Astrophysics}

\vspace*{1.0cm}

Orfeu Bertolami$^{*}$ and Carlos A. D. Zarro$^{**}$\\
\vspace*{0.5cm}
{Departamento de F\'\i sica, Instituto Superior T\'ecnico, \\
Avenida Rovisco Pais 1, 1049-001 Lisboa, Portugal}\\

\vspace*{2.0cm}
\end{center}

\begin{abstract}

\noindent
We consider astrophysical objects such as main-sequence stars, white-dwarfs and neutron stars in a noncommutative context. Noncommutativity is implemented via a deformed dispersion relation $E^{2}=p^{2}c^{2}(1+\lambda E)^{2}+m^{2}c^{4}$ from which we obtain noncommutative corrections to the pressure, particle number and energy densities for radiation and for a degenerate fermion gas. The main implications of noncommutativity for the considered astrophysical objects are examined and discussed.

\end{abstract}

\vfill
\noindent\underline{\hskip 140pt}\\[4pt]
{$^{*}$ Also at Instituto de Plasmas e Fus\~{a}o Nuclear, Instituto Superior T\'ecnico,
Lisboa, Portugal} \\
{E-mail address: orfeu@cosmos.ist.utl.pt} \\
\noindent
{$^{**}$ Also at Instituto de Plasmas e Fus\~{a}o Nuclear, Instituto Superior T\'ecnico,
Lisboa, Portugal} \\
{E-mail address: carlos.zarro@ist.utl.pt}

\newpage


\section{Introduction}

It is believed that noncommutative geometry might play an important role in the description of space-time at scales comparable to Planck length \cite{Connes:1996gi,Madore:1999bi} given that, for instance, noncommutative features arise in string theory in the presence of a constant B field \cite{Witten:1985cc,Seiberg:1999vs}.

Quantum field theory (QFT) in noncommutative spaces has been implemented by substituting the usual pointwise product between fields by the so-called Moyal product. This procedure introduces a minimum length scale, which acts as a UV cut-off even though it is found that IR divergences also appear. Furthermore, in the context of QFT one also encounters problems with causality and unitarity (see, $e.g.$, Refs. \cite{Szabo:2001kg,Douglas:2001ba}). Other issues related to this approach involve the violation of translational invariance \cite{Bertolami:2003nm}, the standing of noncommutative scalar fields on cosmological backgrounds \cite{Lizzi:2002ib,Bertolami:2002eq}, and their stability in curved spaces \cite{Bertolami:2008zv}. Another way to introduce noncommutativity in field theory concerns the generalization of the algebra of noncomutative quantum mechanics \cite{Zhang:2004yu,Bertolami:2005jw,Acatrinei:2003id,Bertolami:2005ud,Bastos:2006kj,Bastos:2006ps} to field algebra \cite{Gamboa:2001fg}. At quantum mechanical level, noncommutativity can be implemented via an extension of the Heisenberg-Weyl algebra and many generalizations of quantum mechanics have been proposed \cite{Zhang:2004yu,Bertolami:2005jw,Acatrinei:2003id,Bertolami:2005ud,Bastos:2006kj,Bastos:2006ps,Gamboa:2001fg,Snyder:1947}. 

Although there are some proposals for a theory of noncommutative gravity  (see, e.g., Refs. \cite{AlvarezGaume:2006bn,Meyer:2005as}), up to now, there is no consistent theory of noncommutative gravity. Thus, rather than considering a noncommutative space over which the fields are defined, one considers instead a deformed dispersion relation for the fields defined in an usual (commutative) space-time \cite{Alexander:2001ck}. This is inspired by the study of quantum groups \cite{AmelinoCamelia:1999pm} and is related to the question of to what extent the Lorentz symmetry is an exact symmetry \cite{Bertolami:2003nm,Bertolami:1999da,Bertolami:2003yi,Bertolami:2003tw} and how special relativity can be modified to accommodate a minimum length scale \cite{Magueijo:2001cr,AmelinoCamelia:2000mn}. This approach has also been considered to address what was believed the puzzle of cosmic rays with energies beyond the Greisein-Zatsepin-Kuzmin cut-off  \cite{Bertolami:2003nm,Bertolami:1999da,Bertolami:2003yi,Bertolami:2003tw} which was later not confirmed observationally \cite{Bergman:2008at,Watson:2008zzb} and to investigate how noncommutativity can account for a inflationary period \cite{Alexander:2001ck,Alexander:2001dr,Barosi:2008gx}. 

The relationship between noncommutativity and a deformed dispersion relation can be understand as sketched in Ref. \cite{AmelinoCamelia:1999pm}. The following commutation relations in the configuration space are considered

\beq\label{eq:1}
[x^{i},t]=i\lambda x^{i}\;\;\;\;\;\;\;\;\;\;\;\;\;\; [x^{i},x^{j}]=0,
\eeq

\noindent where $\lambda$ is constant and $i,j=1,2,3$. Using quantum group methods to define a convenient Fourier transform and a deformed Poincar\'{e} group, one can show that, in momenta variables, the Klein-Gordon operator reads

\beq
\lambda^{-2}(e^{\lambda E}+e^{-\lambda E}-2) - p^{2}c^{2}e^{-\lambda E}=m^{2}c^{4},
\eeq

\noindent which is a deformed dispersion relation.

This can also be seen from the fact that noncommutative theories imply a minimum length, which leads to an extension of special relativity to take into account this new invariant. Defining the boost generator \cite{Magueijo:2001cr}

\beq
K^{i}=L_{0}^{i} + L_{P}p^{i}D
\eeq 

\noindent where $L_{0}^{i}$ is the usual boost generator, $L_{P}$ is the invariant length and $D=p^{\mu}\frac{\partial}{\partial p^{\mu}}$ $ (\mu=0,1,2,3)$ the dilatation generator, it can be seen that the action on momentum space becomes nonlinear and that \cite{Magueijo:2001cr}

\beq
||p||^{2}=m^{2}c^{4}=\frac{p^{\mu}p_{\mu}}{(1-L_{P}E)^{2}}, 
\eeq

\noindent which shows once again that noncommutativity leads to a deformed dispersion relation.

In this work the influence of noncommutativity over astrophysical objects, namely, main-sequence stars, such as the Sun, and more compact objects, such as white dwarfs and neutron stars is investigated. In the present approach one considers the low-energy limit of noncommutativity (cf. below), in opposition to the most common high-temperature limit, which was presumably relevant at the early universe \cite{Alexander:2001ck,Alexander:2001dr}. The aim is to identify the leading noncommutative correction to the thermodynamic quantities used in the description of astrophysical objects.

The considered approach involves a deformed dispersion relation. The use of the grand-canonical formalism of statistical mechanics to obtain the first-order noncommutative corrections to energy, particle number density and pressure. The results are then applied to: standard model of stars (radiation plus nonrelativistic ideal gas) \cite{Chandrasekhar:1967} , white dwarfs (degenerate electron gas) \cite{Chandrasekhar:1967} and neutron stars (Oppenheimer-Volkoff model: degenerate neutron gas) \cite{Oppenheimer:1939ne}. 

This manuscript is organized as follows: in Sec. \ref{sec:defdisprel}, the deformed dispersion relation is presented and some physical features are discussed. In Sec. \ref{sec:defstatmech}, the formalism of the grand-canonical ensemble is employed to compute the leading order noncommutative correction to energy, particle number density and pressure. These results are then applied to obtain noncommutative corrections to radiation, nonrelativistic ideal gas and degenerate fermion gas. Results are used,  in Sec. \ref{sec:aao}, to obtain noncommutative corrections to the main features of astrophysical objects. Finally,  in Sec. \ref{sec:discussions}, one discusses the main physical implications of the obtained results.


\section{Deformed Dispersion Relation}\label{sec:defdisprel}

Inspired in studies of the breaking of Lorentz symmetry at high energies and in theories that admit an invariant length, which as discussed above can be regarded as a noncommutativity expressed by Eqs. (\ref{eq:1}), one defines the deformed dispersion relation, generalizing Ref. \cite{Alexander:2001ck}:

\beq \label{eq:defdisprel}
E^{2}=p^{2}c^{2}(1+\lambda E)^{2} + m^{2}c^{4}.
\eeq

Solving for $E$, one has

\beq \label{eq:completeE}
E=\frac{\lambda p^{2}c^{2} + \sqrt{p^{2}c^{2} + m^{2}c^{4}(1-\lambda^{2}p^{2}c^{2})}}{1-\lambda^{2}p^{2}c^{2}}.
\eeq

Here only the particle branch of the dispersion relation is considered. One then gets four different cases:

\begin{itemize}
\item[(i)] $\lambda p c < 1$, and hence $E>0$.
\item[(ii)] $\lambda p c \rightarrow 1$, and thus $E\rightarrow\infty$.
\item[(iii)] $\lambda p c > 1$ and $|1-\lambda^{2}p^{2}c^{2}|\leq\left(\frac{p}{mc}\right)^{2}$, from which follows that $E<0$. This is a non-physical region for the particle branch.
\item[(iv)] $\lambda p c > 1$ and $|1-\lambda^{2}p^{2}c^{2}|>\left(\frac{p}{mc}\right)^{2}$, which corresponds to a non-physical region since $E$ is complex.
\end{itemize}

\begin{figure}
\centering
\includegraphics{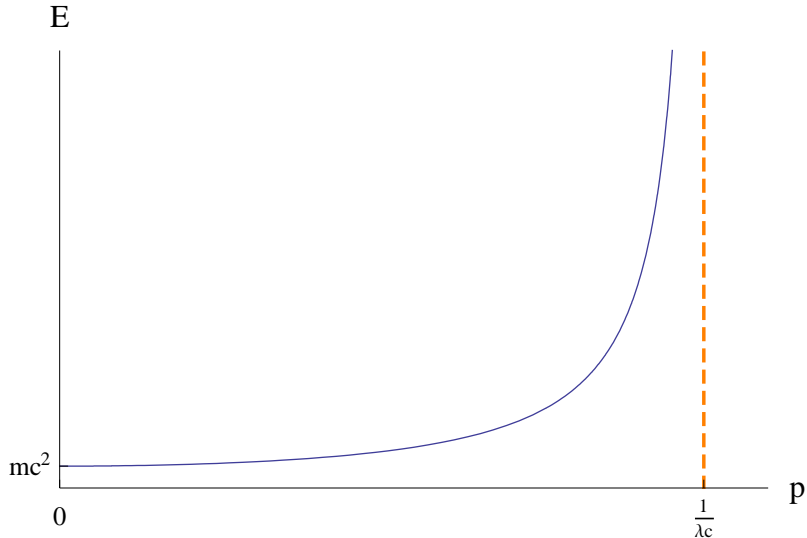}
\caption{Deformed dispersion relation}
\label{fig:defdisprel}
\end{figure}

In Fig. \ref{fig:defdisprel}, the general behavior of this deformed dispersion relation is depicted. The parameter $\lambda$ is associated with the maximum momentum, as although all energies can be attained, one has a maximum momentum,  $p_{max}=1/\lambda c$ \cite{Alexander:2001dr}. 

In this work, since one is interested in astrophysical configurations, and as $\lambda$ is presumably related to the quantum gravity energy scale,  one keeps the first-order correction in $\lambda$. Thus, from Eq. (\ref{eq:completeE}) one obtains

\beq\label{eq:folambda}
E = \lambda p^2c^{2} + \sqrt{p^{2}c^{2}+m^{2}c^{4}}.
\eeq 

Clearly, one recovers the usual relativistic dispersion relation for $\lambda\rightarrow 0$. Notice that the first order correction only acts on the kinetic part of the energy.

\section{Deformed Statistical Mechanics}\label{sec:defstatmech}
Since the noncommutativity is introduced through a deformation of the dispersion relation and the form of the dispersion relation does not alter the foundations of statistical mechanics, one can use the formalism of the grand-canonical ensemble \cite{Alexander:2001ck,Landau:1980,Padmanabhan:2000}.

Consider a system of $N$ particles with energy spectrum given by $\{E_{j}\}$. Each state is labeled by $j$ ($j=1,2,\cdots$) and corresponds to $n_{j}$ particles with energy $E_{j}$. The fugacity is defined as $z=e^{\beta \mu}$, where $\mu$ is the chemical potential and $\beta=(k_{B}T)^{-1}$, being $T$ the temperature and $k_{B}$ the Boltzmann constant. The grand-canonical partition function is obtained from the sum over states

\beq\label{eq:gcpf}
\mathcal{Z}=\sum_{n_{j}}\prod_{j}\left[z e^{-\beta E_{j}} \right]^{n_{j}}=\sum_{n_{j}}\prod_{j}\left[e^{\beta(\mu- E_{j})} \right]^{n_{j}}.
\eeq

The connection to thermodynamics is obtained after introducing the grand-canonical potential defined as

\beq\label{eq:gcpot}
\Phi=-PV=-\frac{1}{\beta}\ln\mathcal{Z},
\eeq

\noindent $P$ being the pressure and $V$ the volume. One can show that \cite{Landau:1980} 

\beq\label{eq:gcpotb}
\Phi=-\frac{1}{a\beta}\sum_{j}\ln(1+aze^{-\beta E_{j}})=-PV,
\eeq

\noindent where the parameter $a$, assumes the following values: $a=1$ for fermions and $a=-1$ for bosons. The average number of particles is given by

\beq\label{eq:anp}
\langle N \rangle=-\left(\frac{\partial \Phi}{\partial \mu}\right)_{\beta}=\sum_{E}n(E),
\eeq

\noindent where $n(E)=\left(z^{-1}e^{\beta E} + a \right)^{-1}$ is the occupation number of particles with energy $E$. The average energy is then obtained

\beq\label{eq:ae}
\langle E \rangle=\sum_{E}E\; n(E)=\sum_{E}\frac{E}{z^{-1}e^{\beta E} + a}.
\eeq

To get the pressure, one considers the large-volume limit: $\sum_{E} \rightarrow \int{\frac{d^{3}\vec{x}d^{3}\vec{p}}{(2\pi\hbar)^{3}}}$. Equation (\ref{eq:gcpotb}) then reads

\beq\label{eq:gcpotc}
\Phi=-PV=-\frac{\gamma}{a\beta}\int{\frac{d^{3}\vec{x}d^{3}\vec{p}}{(2\pi\hbar)^{3}}}\ln(1+aze^{-\beta E(\vec{x},\vec{p})}),
\eeq

\noindent where $E=E(\vec{x},\vec{p})$ and $\gamma$ is the multiplicity of states due to spin. Here one has to consider the nature of the energy function and hence of the deformed dispersion relation. Notice that $E$ in Eq. (\ref{eq:defdisprel}) does not depend on the position and on direction but  only on $p=|\vec{p}|$. Therefore in Eq. (\ref{eq:gcpotc}), the configuration space integration is the trivial and yields the volume $V$. The momentum variable integration yields, after integrating by parts

\beq\label{eq:pressure}
P=\frac{\gamma}{2\pi^{2}\hbar^{3}}\int_{0}^{\frac{1}{\lambda c}}dp \left(\frac{p^{2}}{z^{-1}e^{\beta E} + a} \right)\left(\frac{p}{3}\frac{dE}{dp}\right).
\eeq    

Taking the large-volume limit and integrating over the configuration space, one obtains the particle number density 

\beq\label{eq:numberdensity}
\frac{N}{V}=\frac{\gamma}{2\pi^{2}\hbar^{3}}\int_{0}^{\frac{1}{\lambda c}}dp \left(\frac{p^{2}}{z^{-1}e^{\beta E} + a} \right).
\eeq    

By the same token, the energy density reads

\beq\label{eq:energydensity}
u=\frac{\langle E \rangle}{V}=\frac{\gamma}{2\pi^{2}\hbar^{3}}\int_{0}^{\frac{1}{\lambda c}}dp\;p^{2} \left(\frac{E}{z^{-1}e^{\beta E} + a} \right).
\eeq    

One has then to compute these quantities using Eq. (\ref{eq:defdisprel}) or Eq. (\ref{eq:folambda}) depending on the physical model that describes the  astrophysical objects of interest.

\subsection{Deformed Radiation}
For radiation, a gas of photons, the dispersion relation Eq. (\ref{eq:defdisprel}) is given by 

\beq\label{eq:Erad}
E=pc(1+\lambda E),
\eeq

\noindent which solving for p and changing the integration variable in Eq. (\ref{eq:energydensity}), yields\footnote{For photons $a=-1$, $\mu=0$ and $\gamma=2$.}

\beq\label{eq:uintdef}
u=\frac{1}{\pi^{2}\hbar^{3}c^{3}}\int_{0}^{\infty}dE \frac{E^{3}}{e^{\beta E} - 1}\frac{1}{(1+\lambda E)^{4}}.
\eeq

For $\lambda=0$, one recovers the Stefan-Boltzmann law. This integral cannot be solved analytically. As one is interested in investigating astrophysical objects, one considers the limit $\lambda k_{B} T \ll 1$. This approximation is justified since for the neutron stars, the hottest ones, the central temperature is around ($10^{11}$ $-$ $10^{12}$) K, and to satisfy that condition $\lambda\ll 10 $ GeV$^{-1}$, if $\lambda$ is related to the inverse of the quantum gravity energy scale. The deformed radiation in the limit $\lambda k_{B} T\gg 1$ is considered in Ref. \cite{Alexander:2001ck} to study cosmological issues concerning  the early universe.

Defining a variable $y=\lambda E$, Eq. (\ref{eq:uintdef}) can be written as 

\beq\label{eq:uintdefy}
u=\frac{1}{\pi^{2}\hbar^{3}c^{3}\lambda^{4}}\int_{0}^{\infty}dy \frac{y^{3}}{e^{\frac{y}{\lambda k_{B}T}} - 1}\frac{1}{(1+y)^{4}}.
\eeq

As $\lambda k_{B}T\ll 1$, $(e^{\frac{y}{\lambda k_{B}T}} - 1)^{-1}$ decays exponentially and hence one can expand $y^{3}/(1+y)^{4}$ in Taylor series around zero (any other value of $y$ will be exponentially suppressed). One finds 

\beq\label{eq:uintdefyexp}
u=\frac{1}{\pi^{2}\hbar^{3}c^{3}\lambda^{4}}\left[\int_{0}^{\infty}dy \frac{y^{3}}{e^{\frac{y}{\lambda k_{B}T}} - 1}-4\int_{0}^{\infty}dy \frac{y^{4}}{e^{\frac{y}{\lambda k_{B}T}} - 1}\right].
\eeq

Using the formula {\bf 3.411.1} of Ref. \cite{Gradshteyn:2000} 

\beq\label{eq:formula1}
\int_{0}^{\infty}\frac{x^{\nu -1}\;dx}{e^{\mu x}-1}=\frac{1}{\mu^{\nu}}\Gamma{(\nu)}\zeta{(\nu)} \;\;\;[\mbox{Re}\mu > 0,\mbox{Re}\nu > 0],
\eeq

\noindent where $\Gamma{(\nu)}$ and $\zeta{(\nu)}$ are gamma and zeta functions, respectively. Thus, one obtains the integrals of Eq. (\ref{eq:uintdefyexp}) 

\beq\label{eq:ncsbl}
u=\frac{4\sigma}{c}T^{4} - \frac{96\zeta(5)}{\pi^{2}\hbar^{3} c^{3}}\lambda k_{B}^{5}T^{5},
\eeq

\noindent where $\sigma=\frac{\pi^{2}k_{B}^{4}}{60\hbar^{3} c^{2}}$ is the Stefan-Boltzmann constant. If $\lambda \rightarrow 0$, one recovers the Stefan-Boltzmann law. Equation (\ref{eq:ncsbl}) can be written as: 

\beq\label{eq:ncsblb}
u=\frac{4\sigma_{\mbox{eff}}(T;\lambda)}{c}T^{4},
\eeq 

\noindent where the effective Stefan-Boltzmann constant is given by

\beq
\sigma_{\mbox{eff}}(T;\lambda)=\sigma\left(1-\frac{1440\zeta(5)}{\pi^{4}}\lambda k_{B} T \right).
\eeq

Notice that the first noncommutative correction reduces the energy density. For the pressure, Eq. (\ref{eq:pressure}):

\beq\label{eq:pressurerad}
P=\frac{1}{3\pi^{2}\hbar^{3}c^{3}}\int_{0}^{\infty}dE\;\frac{E^{3}}{e^{\beta E} -1}\frac{1}{(1+\lambda E)^{3}}.
\eeq

Changing the variable to $y=\lambda E$, considering $\lambda k_{B}T \ll 1$ and expanding $\frac{y^{3}}{(1+y)^{3}}$ in Taylor series around zero, one finds

\begin{eqnarray}\label{eq:pressurerada}
P&=&\frac{1}{\pi^{2}\hbar^{3}c^{3}\lambda^{4}}\left[\frac{1}{3}\int_{0}^{\infty}dy \frac{y^{3}}{e^{\frac{y}{\lambda k_{B}T}} - 1}-\int_{0}^{\infty}dy \frac{y^{4}}{e^{\frac{y}{\lambda k_{B}T}} - 1}\right] \nonumber \\
P&=&\frac{4\sigma}{3c}T^{4}-\frac{24\zeta(5)}{\pi^{2}\hbar^{3}c^{3}}\lambda k_{B}^{5}T^{5}
\end{eqnarray}

Dividing Eq. (\ref{eq:pressurerada}) by Eq. (\ref{eq:ncsbl}) one gets

\beq\label{pdividedbyu}
\frac{P}{u}=\frac{1}{3}+(\lambda k_{B}T)\frac{120\zeta(5)}{\pi^{4}},
\eeq

\noindent notice that the usual relationship $u=3P$ is recovered after taking the limit $\lambda\rightarrow 0$. To obtain an equation of state (EOS) $P=P(u)$, one has to write $T$ as a power series of $\lambda$. It suffices to substitute $T=T_{0}+\lambda T_{1}$ in Eq. (\ref{eq:ncsbl}) and compare the terms of the same order in $\lambda$. This yields:

\beq\label{eq:temprad}
T=\left(\frac{15(\hbar c)^{3}}{\pi^{2}}\right)^{1/4}\frac{u^{1/4}}{k_{B}} + \frac{360\zeta(5)(15(\hbar c)^{3})^{1/2}}{\pi^{5}}\frac{\lambda u^{1/2}}{k_{B}}.
\eeq

Substituting Eq. (\ref{eq:temprad}) into Eq. (\ref{pdividedbyu}), one finally gets the first-order noncommutative correction to the EOS

\begin{eqnarray}
P&=&\frac{u}{3}+\frac{120\zeta(5)}{\pi^{4}}\left(\frac{15(\hbar c)^{3}}{\pi^{2}}\right)^{1/4}\lambda u^{5/4}\label{eq:nceosrad}\\
&=&\frac{u}{3}\left[1+\frac{360\zeta(5)}{\pi^{4}}\left(\frac{15}{\pi^{2}}\right)^{1/4}\lambda (\hbar c)^{3/4} u^{1/4}\right]\label{eq:nceosrada}.
\end{eqnarray}

\noindent Equation (\ref{eq:nceosrada}) exhibits a noncommutative correction to the usual EOS of order $\lambda (\hbar c)^{3/4} u^{1/4}$.

\subsection{Nonrelativistic Ideal Gas}
Let us consider now Eq. (\ref{eq:folambda}). In this case one has the first-order noncommutative correction plus the usual dispersion relation. For nonrelativistic ideal gas $p\ll mc$, and Eq. (\ref{eq:folambda}) becomes

\beq
E=\lambda p^{2}c^{2} + mc^2 + \frac{p^{2}}{2m} + \mathcal{O}\left(\frac{p}{mc}\right)^{4}.
\eeq

The noncommutative correction is relevant if the condition $\lambda\;\laq\;(2mc^{2})^{-1}$ is  satisfied. However, for the usual matter of main-sequence stars (hydrogen, for simplicity), this condition implies that $\lambda\;\laq\; 1$ GeV$^{-1}$, which is a too strong restriction to this problem. So no noncommutative correction is considered to the nonrelativistic ideal gas. From the well-known expression for the pressure \cite{Landau:1980},

\beq\label{eq:nrigpressure}
P=\frac{N}{V}k_{B}T,
\eeq

\noindent one can use the definition of the mean molecular weight $\mu_{N}$

\beq\label{eq:mmw}
\mu_{N}=\frac{\rho}{n m_{N}},
\eeq

\noindent where $n=N/V$, $m_{N}$ is the nucleon mass\footnote{$m_{N}=1 \mbox{amu} = 931,494\; \mbox{MeV/c}^{2}$\cite{Amsler:2008zzb}.} and $\rho$ is the mass density to obtain 

\beq\label{eq:nrigpressurefinal}
P=\frac{\rho}{\mu_{N} m_{N}}k_{B}T.
\eeq

\subsection{Deformed Degenerate Fermion Gas} \label{subsec:degfergas}
Consider a low temperature ($T\rightarrow 0$) fermionic system, so that occupation number takes the form $n(E)=H(E_{F} - E)$, where $\mu=E_{F}$ is the Fermi energy above which all levels are not occupied and $H(x)$ is the Heaviside function. The momentum associated to the Fermi energy is the Fermi momentum $p_{F}$. The occupation number written in momentum variable reads $n(p)=H(p_{F}-p)$. As one investigates star configurations, the momenta range in the interval MeV/c $-$ GeV/c, and hence $p_{F}\ll (\lambda c)^{-1}$; that is, one uses the approximate dispersion relation given by Eq. (\ref{eq:folambda}).

The particle number density Eq. (\ref{eq:numberdensity}) for spin one-half particles is given by

\beq \label{eq:numberdensityfermion}
n=\frac{N}{V}=\frac{1}{\pi^{2}\hbar^{3}}\int_{0}^{\frac{1}{\lambda c}}dp n(p) p^{2} = \frac{1}{\pi^{2}\hbar^{3}}\int_{0}^{p_{F}}dp  p^{2} =\frac{p_{F}^{3}}{3\pi^{2}\hbar^{3}}=\frac{(mc)^{3}x^{3}}{3\pi^{2}\hbar^{3}},
\eeq

\noindent where $x=p_{F}/mc$ is defined. The energy density can be computed with the help of Eqs. (\ref{eq:folambda}) and (\ref{eq:energydensity})

\beq\label{eq:energydensityfermion}
u=\frac{E}{V}=\frac{1}{\pi^{2}\hbar^{3}}\int_{0}^{p_{F}}dp\;p^{2}\sqrt{p^{2}c^{2}+m^{2}c^{4}} + \frac{\lambda c^{2}}{\pi^{2}\hbar^{3}}\int_{0}^{p_{F}}dp\;p^{4}.
\eeq

Changing the integration variable to $y=p/mc$, one finds

\beq\label{eq:energydensityfermiona}
u=\frac{E}{V}=\frac{m^{4}c^{5}}{\pi^{2}\hbar^{3}}\int_{0}^{x}dy\;y^{2}\sqrt{y^{2}+1} + \frac{\lambda c^{2}(mc)^{5}}{\pi^{2}\hbar^{3}}\int_{0}^{x}dy\;y^{4}.
\eeq

The first integral can be performed by parts and through the use of  formula {\bf 2.273.3} of Ref. \cite{Gradshteyn:2000}

\beq\label{eq:formula2}
\int_{0}^{x}dy\;\frac{y^{4}}{\sqrt{y^{2}+1}}=\frac{f(x)}{8}=\frac{1}{8}\left(x(2x^{2}-3)\sqrt{1+x^{2}}+3\sinh^{-1}x\right),
\eeq

\noindent where $f(x)=x(2x^{2}-3)\sqrt{1+x^{2}}+3\sinh^{-1}x$ as defined in Ref. \cite{Chandrasekhar:1967}. Finally one obtains the energy density

\begin{eqnarray}
u&=&\frac{(mc^{2})^{4}}{24\pi^{2}(\hbar c)^{3}}\left(8x^{3}\sqrt{1+x^{2}} - f(x) \right) + 	\frac{\lambda (mc^{2})^{5}}{5\pi^{2}(\hbar c)^{3}}x^{5}, \nonumber \\
&=&\frac{(mc^{2})^{4}}{\pi^{2}(\hbar c)^{3}}\left(\frac{x^{3}\sqrt{1+x^{2}}}{3}-\frac{f(x)}{24} +(\lambda mc^{2})\frac{x^{5}}{5}\right).\label{eq:energydensityfermionc}
\end{eqnarray}

\noindent Equation (\ref{eq:energydensityfermionc}) shows that the first noncommutative correction is given by $\lambda mc^{2}$, and since  $mc^{2}$ is in the range  MeV $-$ GeV,  one finds that $\lambda mc^{2}\ll 1$. Taking the limit $\lambda\rightarrow 0$ one recovers the known results  (see, $e.g.$, Ref. \cite{Chandrasekhar:1967}). One computes the pressure using Eqs. (\ref{eq:folambda}) and (\ref{eq:pressure})

\beq\label{eq:pressurefermion}
P=\frac{c^{2}}{3\pi^{2}\hbar^{3}}\int_{0}^{p_{F}}dp\;\frac{p^{4}}{\sqrt{p^{2}c^{2}+m^{2}c^{4}}} + \frac{2\lambda c^{2}}{3\pi^{2}\hbar^{3}}\int_{0}^{p_{F}}dp\;p^{4}.
\eeq 

Following the same procedure to get the energy density, one obtains

\begin{eqnarray}
P&=&\frac{(mc^{2})^{4}}{24\pi^{2}(\hbar c)^{3}}f(x) + 	\frac{2\lambda (mc^{2})^{5}}{15\pi^{2}(\hbar c)^{3}}x^{5}, \nonumber\\
&=&\frac{(mc^{2})^{4}}{\pi^{2}(\hbar c)^{3}}\left(\frac{f(x)}{24} +(\lambda mc^{2})\frac{2x^{5}}{15}\right).\label{eq:pressurefermionc}
\end{eqnarray}


\section{Application to Astrophysical Objects}\label{sec:aao}

\subsection{Main-sequence Stars: The Sun}\label{sec:msssun}

The simplest model of a main-sequence star assumes that a star is composed by a mixture of nonrelativistic ideal gas and radiation maintained in hydrostatic equilibrium by gravity \cite{Chandrasekhar:1967}. The total pressure is $P=P_{\mbox{gas}}+P_{\mbox{rad}}$ where $P_{\mbox{rad}}$ is given by Eq. (\ref{eq:pressurerada}) and $P_{\mbox{gas}}$ is given by Eq. (\ref{eq:nrigpressurefinal}). Defining $\beta_{S}=P_{\mbox{gas}}/P$, or equivalently $P_{\mbox{rad}}=(1-\beta_{S})P$, then $P=P_{\mbox{gas}}/\beta_{S}=	P_{\mbox{rad}}/(1-\beta_{S})$. Substituting Eqs. (\ref{eq:pressurerada}) and (\ref{eq:nrigpressurefinal}) into this relationship one gets

\beq\label{eq:pressureradgas}
\frac{\rho}{\mu_{N} m_{N} \beta_{S}}k_{B}T=\frac{4\sigma}{3c(1-\beta_{S})}T^{4}-\lambda\frac{24\zeta(5)k_{B}^{5}}{\pi^{2}\hbar^{3}c^{3}(1-\beta_{S})}T^5,
\eeq

One rewrites this equation in function of the density $\rho$ for which one substitutes $T=T_{0}+\lambda T_{1}$  and comparing the  powers  in $\lambda$:

\begin{eqnarray}
T_{0}&=&\left[\frac{1-\beta_{S}}{\beta_{S}} \frac{3ck_{B}}{4\sigma \mu_{N} m_{N}} \right]^{1/3}\rho^{1/3},\\
T_{1}&=&\frac{6c\zeta(5)k_{B}^{5}}{\sigma \pi^{2}(\hbar c)^{3}}\left[\frac{1-\beta_{S}}{\beta_{S}} \frac{3ck_{B}}{4\sigma \mu_{N} m_{N}} \right]^{2/3}\rho^{2/3}.
\end{eqnarray}

\noindent Substituting this result into $P_{\mbox{gas}}=\beta P$ and using Eq. (\ref{eq:nrigpressurefinal}) one obtains that $P=K_{1}\rho^{4/3}+\lambda K_{2}\rho^{5/3}$ where

\begin{eqnarray}
K_{1}&=&\frac{(1-\beta_{S})^{1/3}}{(\beta_{S}\mu_{N})^{4/3}}\left(\frac{3ck_{B}^{4}}{4\sigma m_{N}^{4}}\right)^{1/3}=2.67\times 10^{10}\left[\frac{(1-\beta_{S})^{1/3}}{(\beta_{S}\mu_{N})^{4/3}}\right] \;\frac{\mbox{J m}}{\mbox{kg}^{4/3}} \label{eq:k1}\\
K_{2}&=&\frac{(1-\beta_{S})^{2/3}}{(\beta_{S}\mu_{N})^{5/3}}\left(\frac{360\zeta(5)}{\pi^{4}}\right)\left(\frac{3ck_{B}^{4}}{4\sigma m_{N}^{5/2}}\right)^{2/3}=4.52\times 10^{-6}\left[\frac{(1-\beta_{S})^{2/3}}{(\beta_{S}\mu_{N})^{5/3}}\right] \;\frac{\mbox{J$^{2}$ m$^{2}$}}{\mbox{kg}^{5/3}}\label{eq:k2}.
\end{eqnarray}

If one assumes that $\beta_{S}$ is constant over all stars and that its chemical composition is unchanged so that $\mu_{N}$ is also constant\footnote{This is know as Eddington's standard model of stars \cite{Padmanabhan:2001}.}, one finds that the matter is a polytrope with polytropic index $n=3$, perturbed by a polytrope with $n=3/2$. The question of stability can be roughly analyzed in the following terms: the dominant term is $\Gamma=4/3$ where $\Gamma$ is defined as $\Gamma=1-\frac{1}{n}$. This represents the point where a configuration is marginally stable and one shows that this configuration is stable if $\Gamma>\frac{4}{3}+\frac{2.25 GM}{c^{2}R}$ after general relativity corrections \cite{Padmanabhan:2001}. So the latter correction has the effect to decrease the region of stability of this configuration. The noncommutative correction to pressure does not affect the problem of stability since it yields $\Gamma>\frac{5}{3}$ and a positive contribution. Indeed, following Ref. \cite{Padmanabhan:2001}, neglecting general relativity corrections,  the energy of this configuration comprises the sum of the internal energy, which is proportional to $PV$, and the gravitational potential energy, proportional to $\frac{GM^{2}}{R}$. The pressure is given by $P=K_{1}\rho^{\Gamma_{1}}+\lambda K_{2}\rho^{\Gamma_{2}}$ and $\lambda\frac{K_{2}}{K_{1}}\rho^{\Gamma_{2}-\Gamma_{1}}\ll 1$. The energy then reads

\beq
E =k_{0}PV-k_{1}\frac{M^{2}}{R} =C_{1}M\rho_{c}^{\Gamma_{1}-1}+\lambda C_{2}M\rho_{c}^{\Gamma_{2}-1}-k_{3}M^{5/3}\rho_{c}^{1/3},
\eeq

\noindent where $k_{0},k_{1},k_{3},C_{1},C_{2}$ are constants and $\rho_{c}$ is the central density. Calculating the value of $M$ at the critical point $\frac{\partial E}{\partial \rho_{c}}=0$, one obtains

\beq \label{eq:mcrit}
M = \left[\frac{3}{k_{3}}\left(C_{1}(\Gamma_{1}-1)\rho_{c}^{\Gamma_{1}-4/3} +\lambda C_{2}(\Gamma_{2}-1)\rho_{c}^{\Gamma_{2}-4/3} \right) \right]^{3/2}.
\eeq

A necessary, but not sufficient, condition for stability of the configuration is $\frac{d\ln M}{d\ln \rho_{c}}>0$, for $M$ given by $\frac{\partial E}{\partial \rho_{c}}=0$ \cite{Glendenning:1997wn}. Using Eq. (\ref{eq:mcrit}), one gets

\beq
\frac{d\ln M}{d\ln \rho_{c}}=\frac{3}{2}\left(\Gamma_{1}-\frac{4}{3}\right)+\lambda\frac{3C_{2}(\Gamma_{2}-1)}{2C_{1}(\Gamma_{1}-1)}(\Gamma_{2}-\Gamma_{1})\rho_{c}^{\Gamma_{2}-\Gamma_{1}}.
\eeq

In the model investigated here, $\Gamma_{1}=4/3$ and $\Gamma_{2}=5/3>\Gamma_{1}$, so $\frac{d\ln M}{d\ln \rho_{c}}>0$. Hence the effect of noncommutativity is to turn a marginal stable configuration into a stable one. 

As an example, one applies this formalism to the Sun.  At the center, $1-\beta_{S} \approx 10^{-3}$, $\rho_{c}=1.53\times10^{5}$ kg/m$^{3}$ and $\mu_{N}=0.829$ \cite{Padmanabhan:2001}; hence, one can estimate the relevance of the noncommutative correction over the pressure, 

\beq \label{eq:ratiopncpcmss}
\frac{P_{NC}}{P_{C}}=\frac{\lambda K_{2}\rho^{5/3}}{K_{1}\rho^{4/3}}=  9.66\times10^{-16}\lambda,
\eeq

\noindent where $P_{NC}=\lambda K_{2}\rho^{5/3}$ is the noncommutative correction to the pressure, $P_{C}=K_{1}\rho^{4/3}$ is the usual (commutative) term and $\lambda$ must be expressed in J$^{-1}$ units. One computed this value at the center of the star using Eqs. (\ref{eq:k1}) and (\ref{eq:k2}). A discussion including the value of $\lambda$ will be postponed until Sec. \ref{sec:discussions}, but one cannot fail to see that this correction is, as expected, quite small. Notice that given that the Sun is a fairly typical main-sequence star of its class, the obtained results can be seen as quite general.  

\subsection{White Dwarfs}
In last subsection, one has seen that the noncommutative correction to main-sequence stars is very small. Since the noncommutative corrections are supposed to be relevant for configurations with higher energy per particle, one considers next white-dwarfs. Astronomical data \cite{Bhatia:2001} show that these stars have mass of the order of the Sun and  planet sizes, hence the range of mass density at the center  is $10^{8}$ kg/m$^{3}$ $\laq\; \rho_{c}\;\laq\;10^{12}$ kg/m$^{3}$. This requires considering the quantum behavior of  matter. Given that all  atoms are ionized
and so  electrons are free, the physical assumption is that it is the pressure of this electron gas that balances the gravitational force. Hence the appropriate formalism to investigate noncommutative correction is the one developed in Sec. \ref{subsec:degfergas}.  The degeneracy of this electron gas is justified since the temperature associated to $E_{F}$ is greater than the temperature of the white dwarfs\footnote{As noted in Ref. \cite{Padmanabhan:2001} this approximation is valid inside the star but not in its envelope; however this discussion can be skipped since it is only relevant in the study of thermal evolution of these stars, which is beyond the scope of this work.}. Indeed, suppose that $E_{F}\sim1$ MeV, so $T\sim10^{10}$ K which is a much larger than the usual internal temperature of these stars ($T\sim10^{7}$ K).

One defines the electronic molecular weight as 

\beq\label{eq:elmolweight}
\mu_{e}= \frac{\rho}{n_{e}m_{p}},
\eeq

\noindent where $n_{e}$ is the electronic particle number density and $m_{p}$ is the proton mass. Substituting Eq. (\ref{eq:elmolweight}) into Eq. (\ref{eq:numberdensityfermion}), one finds 

\beq\label{eq:xwd}
x=(3\pi^{2})^{1/3}\frac{\hbar}{m_{e}c}n_{e}^{1/3}=\left(\frac{3\pi^{2}}{m_{p}}\right)^{1/3}\frac{\hbar}{m_{e}c}\left(\frac{\rho}{\mu_{e}}\right)^{1/3}=10^{-3}\left(\frac{\rho}{\mu_{e}}\right)^{1/3},
\eeq

\noindent where $m_{e}$ is the mass of the electron and this result is presented in SI units.  Computing now the value of $x$ for $\rho=10^{8}$ kg/m$^{3}$, the lowest white-dwarfs density, $\rho=2\times10^{9}$ kg/m$^{3}$ (region where $p_{F}\sim m_{e}c$) and $\rho=10^{12}$ kg/m$^{3}$, the largest white-dwarf density, and $\mu_{e}=2$, which corresponds to a star formed by helium, one obtains, respectively,

\begin{eqnarray}
x_{8}&=&0.37 \label{eq:xnotsodense} \\
x_{9}&=&1.01  \label{eq:xintermediate} \\
x_{12} &=& 7.99 \label{eq:xdense}.
\end{eqnarray}

To quantify the relevance of noncommutativity,  one  divides the first-order noncommutative correction to pressure ($P_{NC}$) by its commutative part ($P_{C}$) [cf. Eq. (\ref{eq:pressurefermionc})]:

\beq\label{eq:pertpressurewd}
\frac{P_{NC}}{P_{C}}=\frac{48x^{5}}{15 f(x)} \lambda m_{e}c^{2}.
\eeq

\begin{figure}
\centering
\includegraphics{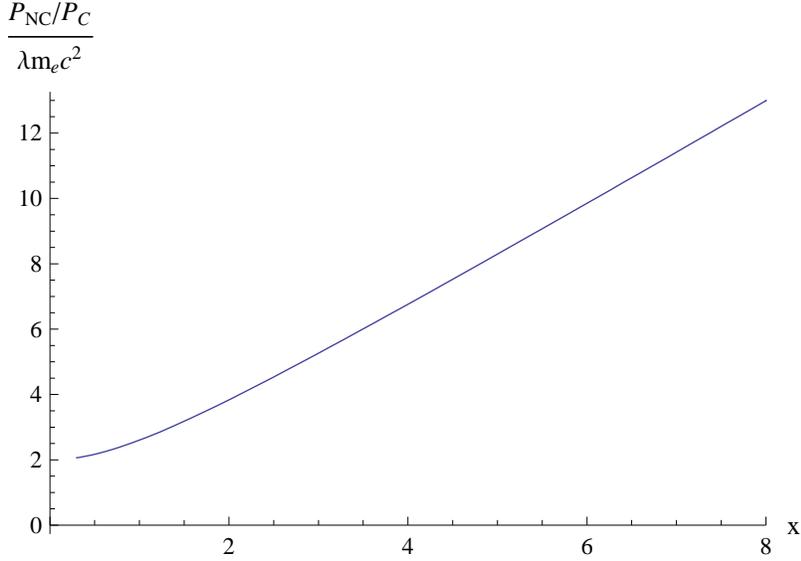}
\caption{Ratio between the noncommutative part and the commutative part of the pressure in units of $\lambda m_{e}c^{2}$}
\label{fig:fermi}
\end{figure}

In Fig. \ref{fig:fermi}, it is plotted $\frac{P{NC}}{P_{C}}$ against $x$, for $0.3\;\laq\; x\; \laq\; 8$, usual values found in white-dwarfs. Notice this ratio has the order of only few $\lambda m_{e} c^{2}$, as $m_{e} c^{2}\sim0.5$ MeV and $\lambda$ is probably much smaller than this quantity (see Sec. \ref{sec:discussions}). Notice that other corrections have been neglected, such as the ones due to general relativity, Coulomb interaction at low densities, high density matter and even corrections due to rotation and magnetic field, which can presumably be much larger than the corrections due to noncommutativity. 
 
Finally, in order to tackle the problem of stability, one  has to proceed as in Sec. \ref{sec:msssun}.  One skips the full derivation (see, $e.g.$, Ref. \cite{Padmanabhan:2001}), but consider instead a quick analysis. The noncommutative term of Eq. (\ref{eq:pressurefermionc}), it is proportional to $x^{5} \propto \rho^{5/3}$ [Eq. (\ref{eq:xwd})], and this corresponds to a polytrope for which $\Gamma=5/3$, that lies in the range of stability. So the noncommutative correction does not introduce any instability to white dwarfs.

\subsection{Neutron Stars}
For higher densities ($\rho \;\gaq\; 10^{12}$ kg/m$^{3}$), the stars become richer in neutrons that are formed by the combination of electrons and protons. These stars have masses comparable to the Sun ($M\sim1 M_{\odot}$) and radius about $R\sim 10$ km; thus due to this very nature, these objects must be described by general relativity, given that $\frac{GM}{Rc^{2}}\sim0.1$. 

One considers the simplest model to describe these stars, namely a degenerate neutron ideal gas. This is the well-known Oppenheimer-Volkoff (OV) model \cite{Oppenheimer:1939ne}. Although, an ideal gas approximation is unrealistic, since it is necessary to consider the existence of nuclear forces, this  is a good starting point. One first  obtains a thermodynamic description  (see Sec. \ref{subsec:degfergas}). One finds that in this model the  EOS that takes in account the interaction between the neutrons is of the same order of magnitude of the OV approximation \cite{Glendenning:1997wn}. Considering perturbative calculations up to second order in the strong coupling of QCD for a  cold quark model shows that the interaction reduces the pressure of the ideal gas model \cite{Fraga:2001xc}. So the ideal gas model is stiffer than an EOS that includes nuclear interaction. By simplicity, the Oppenheimer-Volkoff model will be used here. Neutron stars are formed at $10^{12}$ K and then, due to neutrino emission, they quickly attain $10^{9}$ K ($\sim 10^{-1}$ MeV), which is much lower than the corresponding Fermi energy of neutrons ($\sim 1$ GeV); this indicates that the neutron gas is degenerate.

Using Eqs. (\ref{eq:numberdensityfermion}), (\ref{eq:energydensityfermionc}) and (\ref{eq:pressurefermionc}), one finds

\begin{eqnarray}
x&=&\frac{\hbar}{m_{n}c}\left(3\pi^{2} n\right)^{1/3}, \label{eq:xneutron} \\
u&=&\frac{(m_{n}c^{2})^{4}}{\pi^{2}(\hbar c)^{3}}\left(\frac{x^{3}\sqrt{1+x^{2}}}{3}-\frac{f(x)}{24} +(\lambda m_{n}c^{2})\frac{x^{5}}{5}\right),\label{eq:energydensityneutron} \\
P&=&\frac{(m_{n}c^{2})^{4}}{\pi^{2}(\hbar c)^{3}}\left(\frac{f(x)}{24} +(\lambda m_{n}c^{2})\frac{2x^{5}}{15}\right),\label{eq:pressureneutron}
\end{eqnarray}

\noindent where $m_{n}$ is the neutron mass. In order to obtain an EOS $P=P(u)$, one has to solve Eq. (\ref{eq:energydensityneutron}) for $x$ and substitute it into Eq. (\ref{eq:pressureneutron}). This can  be done numerically. Since one is interested in getting the order of magnitude of the perturbation, one should perform the ratio between this correction and the commutative result. For  usual baryonic matter, $n=0.15\times 10^{45}$ m$^{-3}$ \cite{Glendenning:1997wn}, one gets $x=0.35$ and substituting this into Eq. (\ref{eq:pressureneutron}), one obtains that $\frac{P_{NC}}{P_{C}}=2.1(\lambda m_{n} c^{2})$. Thus for typical $x$ values, the results do not differ from the white-dwarf case: Therefore, the results shown in Fig. \ref{fig:fermi} remain valid, after performing $m_{e}\rightarrow m_{n}$. Hence the effect of noncommutativity is a few $\lambda m_{n}c^{2}$.  The effect of noncommutativity is more relevant for denser configurations. 

The issue of stability for neutron stars is more evolved than the stability of main-sequence stars and white dwarfs given that the description of  neutron stars is not Newtonian. The Oppenheimer-Volkoff equation must be solved for $u$ and $P$ [Eqs. (\ref{eq:energydensityneutron}) and (\ref{eq:pressureneutron})], from where  the stability must be analyzed. This will be examined elsewhere \cite{Bertolami:2009}.  


\section{Discussions and conclusions}\label{sec:discussions}
In this work one has considered the effects of noncommutativity in astrophysical objects. Noncommutativity is implemented via a deformed dispersion relation and its implication for the thermodynamical quantities that are computed through the grand-canonical ensemble formalism. One has examined main-sequence stars, white dwarfs, and neutron stars.

Up to TeV scale, there is no experimental evidence of a deformed dispersion relation \cite{Amsler:2008zzb}; hence, a good starting point would have to consider the value of $\lambda \ll 10^{-3}$ (GeV)$^{-1}$. The value of $\lambda$ can be obtained from ultrahigh energy cosmic rays experiments or by quantum gravity arguments. The most stringent limit to this quantity is  $\lambda<2.5\times 10^{-19}$ GeV$^{-1}$ $=1.6\times10^{-9}$ J$^{-1}$ \cite{Bertolami:2003yi}.  It justifies one to consider only up to the first-order correction due to noncommutativity. Noncommutative correction to relevant statistical mechanical quantities is obtained and these results were applied in simple models to describe three different types of stars.  

For main-sequence stars one gets for the ratio of noncommutative correction and the leading term for pressure $\frac{P_{NC}}{P_{C}}\sim 10^{-24}$, for $\lambda\sim 10^{-19}$ GeV$^{-1}$. Furthermore, it is also shown that the noncommutative correction moves the stability region of these stars towards a more stable situation. For white dwarfs, the relevance of this correction is of order $\lambda m_{e}c^{2}$, as $m_{e}c^{2}=0.5$ MeV, one obtains $\frac{P_{NC}}{P_{C}}\sim 10^{-22}$, and for neutron stars $\frac{P_{NC}}{P_{C}}\sim 10^{-19}$ for $\lambda\sim 10^{-19}$ GeV$^{-1}$. Thus, one finds that the effects of noncommutativity are increasing important from main-sequence stars to neutron stars. This leads one to believe that if the space is noncommutative it might have a relevant impact on black-hole physics. Indeed, phase-space noncommutativity effects are shown to play an important role on the thermodynamics of Schwarzschild black holes \cite{Bastos:2009ae}.

Finally, one finds that likewise for the case of main-sequence stars, the noncommutative correction turns, however small, white-dwarfs more stable. For neutron stars, however a proper analysis has still to be performed.


\section*{Acknowledgments}


The work of C. A. D. Z. is fully supported by the FCT SFRH/BD/29446/2006. The authors would like to thank Jorge P\'{a}ramos and Bruno Mintz for some enlightening comments.

\newpage


\appendix
\newpage
\vfill

\end{document}